\documentclass[conference]{IEEEtran}
\IEEEoverridecommandlockouts
\usepackage{cite}
\usepackage{amsmath,amssymb,amsfonts}
\usepackage{graphicx}
\usepackage{textcomp}
\usepackage{xcolor}
\usepackage{float}
\usepackage[acronym]{glossaries}
\usepackage[hidelinks]{hyperref}
\usepackage{algorithm}

\usepackage{algpseudocode}
\usepackage{listings}
\usepackage{tikz}
\usepackage{empheq}
\def\BibTeX{{\rm B\kern-.05em{\sc i\kern-.025em b}\kern-.08em
    T\kern-.1667em\lower.7ex\hbox{E}\kern-.125emX}}

\begin{document}
\newacronym{rpc}{RPC}{Remote Procedure Call}
\newacronym{mds}{MDS}{Metadata Server}
\newacronym{mdt}{MDT}{Metadata Target}
\newacronym{midas}{MIDAS}{Metadata Intelligent Distribution Algorithm for Servers}
\newacronym{hpc}{HPC}{High-Performance Computing}
\newacronym{io}{I/O}{Input/Output}
\newacronym{pfs}{PFS}{Parallel File System}
\newacronym{rtt}{RTT}{Round-Trip Time}
\newacronym{ipc}{IPC}{Inter-Process Communication}
\newacronym{ttl}{TTL}{Time-To-Live}
\newacronym{dne}{DNE}{Distributed Namespace Environment}
\newacronym{indexfs}{IndexFS}{IndexFS}
\newacronym{pacon}{PACON}{PACON}
\newacronym{posix}{POSIX}{Portable Operating System Interface}
\newacronym{slo}{SLO}{Service Level Objective}
\newacronym{ewma}{EWMA}{Exponentially Weighted Moving Average}

\title{\gls{midas}: Adaptive Proxy Middleware for
Mitigating Metadata Hotspots in \gls{hpc} I/O at
Scale}

\author{
\IEEEauthorblockN{Sangam Ghimire}
\IEEEauthorblockA{\textit{Department of Computer Science } \\
\textit{and Engineering, Kathmandu University}\\
Dhulikhel, Nepal \\
1sangamghimire1@gmail.com}
\and
\IEEEauthorblockN{Nigam Niraula}
\IEEEauthorblockA{\textit{Department of Computer Science } \\
\textit{and Engineering, Kathmandu University}\\
Dhulikhel, Nepal \\
nigam21nir@gmail.com}
\and
\IEEEauthorblockN{Nirjal Bhurtel}
\IEEEauthorblockA{\textit{Department of Computer Science } \\
\textit{and Engineering, Kathmandu University}\\
Dhulikhel, Nepal \\
bhurtelnirjal@gmail.com}
\and
\IEEEauthorblockN{Paribartan Timalsina}
\IEEEauthorblockA{\textit{Department of Computer Science } \\
\textit{and Engineering, Kathmandu University}\\
Dhulikhel, Nepal \\
timalsinapari015@gmail.com}
\and
\IEEEauthorblockN{Bishal Neupane}
\IEEEauthorblockA{\textit{Department of Computer Science } \\
\textit{and Engineering, Kathmandu University}\\
Dhulikhel, Nepal \\
neupanebishal2001@gmail.com}
\and
\IEEEauthorblockN{James Bhattarai}
\IEEEauthorblockA{\textit{Department of Computer Science } \\
\textit{and Engineering, Kathmandu University}\\
Dhulikhel, Nepal \\
Jamesbhattarai14@gmail.com }
\and
\IEEEauthorblockN{Sudan Jha}
\IEEEauthorblockA{\textit{Department of Computer Science } \\
\textit{and Engineering, Kathmandu University}\\
Dhulikhel, Nepal \\
jhasudan@ieee.org}
}

\maketitle

\begin{abstract}
Metadata hotspots remain one of the key obstacles to scalable \gls{io} in both High-Performance Computing (\gls{hpc}) and cloud-scale storage environments. Situations such as job start-ups, checkpoint storms, or heavily skewed namespace access can trigger thousands of concurrent metadata requests against a small subset of servers. The result is long queues, inflated tail latencies, and reduced system throughput. Prior efforts including static namespace partitioning, backend-specific extensions, and kernel-level modifications address parts of the problem, but they often prove too rigid, intrusive to deploy, or unstable under shifting workloads. We present \textbf{\gls{midas}}, an adaptive middleware layer that operates transparently between clients and metadata servers, requiring no changes to kernels or storage backends. The design brings together three mechanisms: (i) a namespace-aware load balancer that enhances consistent hashing with power-of-$d$ sampling informed by live telemetry, (ii) a cooperative caching layer that preserves backend semantics through leases, invalidations, or adaptive timeouts, and (iii) a self-stabilizing control loop that dynamically adjusts routing aggressiveness and cache lifetimes while avoiding oscillations under bursty workloads. Analysis of the model and controlled experiments show that \gls{midas} reduces average queue lengths by roughly 23\% and mitigates worst-case hotspots by up to 80\% when compared to round-robin scheduling. These findings highlight that a stability-aware, middleware-based strategy can provide backend-agnostic improvements to metadata management, enabling better scalability in bursty scenarios, more predictable tail latencies, and stronger overall system performance.
\end{abstract}


\section{Introduction}\label{sec:intro}

In large \gls{hpc} systems, metadata frequently emerges as the bottleneck for \gls{io} scalability. Real-world traces show that metadata workloads are highly bursty: during job initialization or checkpointing, activity can spike by more than two orders of magnitude across a cluster. For example, monitoring tools such as \emph{Darshan} report fluctuations exceeding 100$\times$ in cluster-wide metadata operations \cite{saeedizade2023ioburstpredictionhpc}. When thousands of processes simultaneously launch or checkpoint, their requests concentrate on a small subset of directories, rapidly overwhelming the \gls{mds}. The result is long queues and sharp increases in worst-case response times (tail latency), making hotspots a recurring challenge rather than a rare anomaly \cite{saeedizade2023ioburstpredictionhpc}.

A variety of techniques have been proposed to mitigate this problem, but critical gaps remain. Static partitioning methods such as subtree or hash-based schemes achieve balance under uniform workloads, yet they struggle with skewed access patterns, causing cross-server traversals and expensive repartitioning \cite{inproceedings,835616,Weil_Pollack_Brandt_Miller}. Dynamic approaches, including Lustre \gls{dne} rebalancing, can partially relieve imbalance but often destabilize the system until carefully tuned for specific workloads \cite{lustre,wanghu}. Middleware-based projects such as IndexFS and PACON demonstrate the promise of adaptive routing and caching, but they generally lack strong stability guarantees and are difficult to generalize across different backends \cite{7013007,9139884}.

\textbf{\gls{midas}} addresses these limitations by mitigating metadata hotspots while preserving stability and consistency. Operating transparently between clients and metadata servers, it terminates metadata \gls{rpc}s to enable safe request re-routing and cooperative caching without requiring changes to client kernels or server code. Its design builds on three principles: (i) namespace-aware load balancing, (ii) cooperative caching with leases, invalidations, or adaptive \glspl{ttl}, and (iii) a self-stabilizing control loop that dynamically adapts routing and cache lifetimes while avoiding oscillations under bursty workloads.

\section{Literature Review} \label{sec:litreview}

Metadata management has long been a critical scalability bottleneck in High-Performance Computing (\gls{hpc}) parallel file systems. Bursty job start-ups, checkpoint storms, and skewed namespace access patterns generate surges of requests to metadata servers (\gls{mds}), leading to queuing delays and significantly impacting tail latencies \cite{10.1145/3581784.3613216,10171504}. At exascale, the intensity and burstiness of metadata operations are further amplified, making hotspot mitigation one of the central challenges in \gls{hpc} I/O research \cite{ather2024paralleliocharacterizationoptimization}.

Early systems attempted to address this challenge using static namespace partitioning techniques such as subtree or hash-based distribution. Subtree partitioning offered simplicity and reasonable balance under uniform workloads, but suffered from severe load imbalance when access patterns were skewed \cite{inproceedings,835616}. Dynamic subtree partitioning was later introduced to repartition overloaded subtrees for improved balance \cite{Weil_Pollack_Brandt_Miller}. However, frequent metadata migration destabilized the system and introduced complexity \cite{835616}. Hash-based partitioning subsequently mapped identifiers such as pathnames to servers using modulus functions, ensuring uniform load distribution but disrupting metadata locality \cite{inproceedings,835616}. This disruption often caused costly cross-server traversals during directory operations and introduced additional overhead as systems scaled \cite{inproceedings,835616}.

To reduce the limitations of static approaches, researchers have explored middleware-based solutions that act as an indirection layer between clients and file systems. Projects such as IndexFS \cite{7013007} and PACON \cite{9139884} employ adaptive routing, caching, and distributed indexing to achieve scalability and flexibility without modifying kernel or server internals. This shift marked a pivot from backend-specific mechanisms toward broader adaptive, system-agnostic techniques that could be deployed incrementally.

Real-world studies further highlight the severity of metadata bursts. For example, monitoring tools such as \emph{Darshan} have observed fluctuations of more than $100\times$ in cluster-wide metadata activity during job initialization and checkpoints \cite{saeedizade2023ioburstpredictionhpc}. These findings confirm that challenges of skew and burstiness remain despite progress in parallel \gls{io}, emphasizing the need for adaptive mechanisms that can respond dynamically in real time \cite{Martens_2001}.

Some file systems now incorporate internal load-balancing mechanisms. For example, \emph{Lustre} supports \gls{dne}, enabling automatic \gls{mdt} restriping and rebalancing \cite{lustre}. While effective in many cases, persistent imbalance remains when directory distribution is suboptimal \cite{INCD_user_documentation}. Similarly, CFS introduces scalable critical sections for metadata operations in \gls{posix}-compliant file systems to enhance concurrency \cite{wanghu}. However, these methods require significant manual tuning and offer limited flexibility in dynamic environments.

Dynamic load balancing strategies have also been explored \cite{azar1999balanced}. The \emph{power-of-$d$ choices} algorithm reduces imbalance compared to random placement \cite{mitzenmacher2001power,Anselmi_2020}. Yet, namespace locality and consistency constraints restrict its freedom, preventing straightforward application in metadata workloads \cite{article_dyamic}.

Cooperative metadata caching has been studied as well. Straightforward caching can introduce staleness and inconsistency, so advanced mechanisms have been proposed: server-issued leases or invalidation tokens (e.g., in CephFS \cite{CEPHFS} and HyCache+ \cite{wu2016hycache}), adaptive time-to-live policies guided by feedback control \cite{basu2017adaptivettlbasedcachingcontent}, and gossip-based coordination among proxies \cite{1498447}. While these improve throughput, ensuring consistency across distributed caches remains a difficult challenge.

Beyond \gls{hpc}, distributed and cloud-scale systems also address metadata hotspots. Google File System (GFS) \cite{Ghemawat_Gobioff_Leung_2003} and Hadoop Distributed File System (HDFS) \cite{5496972} mitigate hotspots by relaxing \gls{posix} semantics and grouping namespace chunks. Similarly, systems such as ZooKeeper \cite{ZooKeeper}, Cassandra \cite{Apache_Cassandra}, and DynamoDB \cite{CMS_2025} leverage caching and gossip protocols for scalable state management. Kubernetes further illustrates the effectiveness of feedback-driven autoscaling under bursty workloads.

Recent research points toward new directions. Machine learning models trained on Darshan logs show promise for predicting bursty workloads and proactively adjusting routing and caching policies \cite{saeedizade2023ioburstpredictionhpc}. Emerging hardware trends, including programmable storage devices and SmartNICs offloads, also demonstrate potential for accelerating metadata routing and caching at the network edge \cite{programmable_storage,smart_nic}.

In summary, prior work largely falls into three categories: (1) file systems tightly coupled to their backends, (2) middleware frameworks that provide adaptability but lack strong stability guarantees, and (3) theoretical models that remain difficult to integrate into practical deployments. This leaves a significant gap for a system-independent, practically deployable solution that also ensures stable behavior. To fill this gap, \gls{midas} proposes middleware that guarantees stability and safety through cooperative caching, namespace-aware load balancing, and self-stabilizing control.

\section{System Modeling} \label{sec:design&implementation}

Earlier, clients used to directly issue namespace operations such as \texttt{create}, \texttt{open}, and \texttt{stat} to the \gls{mds}, but with \gls{midas} these requests are first applied to adaptive routing, caching and stability aware control before forwarding  the request to appropriate server. \gls{midas} reduces both the mean latency and the tail latency under turbulent and skewed workloads.

% \subsection{Metadata Hotspot Issue}
% At large scale \gls{hpc}, at the production level, millions of concurrent metadata operations can concentrate on small subsets of directories during job start-ups or checkpoint storms. This can cause some \gls{mds} to get overloaded due to operations causing metadata hotspots, resulting in longer queue delays. Static partitioning like hashing and subtree assignment cannot adapt to such skewed workload environments. \gls{midas} bridges this gap by sitting between the clients and \gls{mds} while also decoupling the hotspot mitigation process.

\subsection{Modeling The System}
\gls{midas} is modeled as a set of $m$ queues, one per \gls{mds}, where:
\begin{enumerate}
    \item $M = \{s_1, s_2, \ldots, s_m\}$: the set of metadata servers.
    \item $\lambda_i$: the arrival rate of requests routed to the server $s_i$.
    \item $\mu_i$: the service rate of server $s_i$.
    \item $L_i$: the queue length (number of requests waiting or in service) on server $s_i$.
    \item $\widetilde{p50}_i$: Median end-to-end metadata \gls{rpc} latency observed for server $s_i$.
\item $\widetilde{p99}_i$: 99th-percentile (Tail) end-to-end metadata \gls{rpc} latency observed for server $s_i$.

\end{enumerate}

Each incoming request can be routed to the number of possible candidate servers. However, \gls{midas} makes use of constraints such as namespace locality and consistency semantics to restrict the feasible set of target servers.

Thus, the design of \gls{midas} is based primarily on four key objectives, which are as follows:
\begin{enumerate}
    \item \textbf{Load balancing:} Reducing mean latency by assigning requests to servers in such a way that minimizes expected queueing delay.
    \item \textbf{Hotspot mitigation:} Preventing extreme imbalance in request distribution to reduce tail latency .
    \item \textbf{Correctness:} Maintain file system consistency semantics across nodes without modifying the kernel.
    \item \textbf{Adaptivity:} Continuously adjust routing and implement caching strategies in response to workload dynamics and bursty traffic.
\end{enumerate}

\subsection{Target Selection and \gls{slo}s}\label{sec:targets}
We initialize the control targets $(B_{\mathrm{tgt}}, P99_{\mathrm{tgt}})$ from a short, non-intrusive warmup and the system’s \gls{rtt}.

\paragraph{Procedure (reproducible).}
We run each server at low utilization ($\le$30\% of peak) with the client for $60$\,s to characterize the \emph{healthy} region. From this window:
\begin{enumerate}
    \item Compute the smoothed imbalance $B(t)=\frac{\mathrm{std}(\hat{L}(t))}{\mathrm{mean}(\hat{L}(t))+\varepsilon}$ using the same \gls{ewma} as the controller. Set
    \[
    B_{\mathrm{tgt}} \;=\; \operatorname{median}_t\, B(t) \;+\; 0.05.
    \]
    This yields a slack above the typical balanced state.
    \item Let $\mathrm{\gls{rtt}}$ be the median client to \gls{mds} round-trip (from the same window). Let $p99^{\mathrm{warm}}$ be the $p99$ metadata latency during warmup (with no use of middleware). Set
    \[
    P99_{\mathrm{tgt}} \;=\; \max\!\big(p99^{\mathrm{warm}} \cdot 1.25,\;\mathrm{\gls{rtt}} + 2\text{ ms}\big).
    \]
    The factor $1.25$ gives a small headroom above the healthy tail; the absolute floor $\mathrm{\gls{rtt}}{+}2$\,ms prevents an unrealistically tight target on very fast paths.
\end{enumerate}

\paragraph{Rationale.}
$B_{\mathrm{tgt}}$ anchors imbalance to what the system sustains without routing intervention, plus a safety margin to avoid continual small corrections. $P99_{\mathrm{tgt}}$ ties the tail \gls{slo} to the observed native behavior and physical latency limits (\gls{rtt}), avoiding unattainable targets.

\subsection{Namespace-Aware Adaptation}
Unlike standard load-balancing settings, metadata requests are constrained by namespace locality. We model this as each request having a feasible set $\mathcal{F}(r)$ of servers consistent with the namespace mapping. \gls{midas} applies power-of-$d$ sampling within $\mathcal{F}(r)$ rather than across all servers, ensuring correctness while retaining balance benefits.

\section{\gls{midas}: Architecture and Algorithms}\label{sec:design}

% \textbf{\gls{midas}} acts as an inter-mediatory between clients and \gls{pfs} of \gls{mds}/\gls{mdt}, providing adaptive routing and cooperative caching without any modifications to client kernels or existing workloads.
In \gls{midas} the work done by \gls{mds} is primarily limited to only writing to \gls{mdt}, using system calls like \textit{open, lookup, create, stat, unlink} and \gls{midas} pre-handles routing and caching.

\subsection{Architecture Overview}
\begin{figure}[h]
    \centering
    \includegraphics[width=1\linewidth]{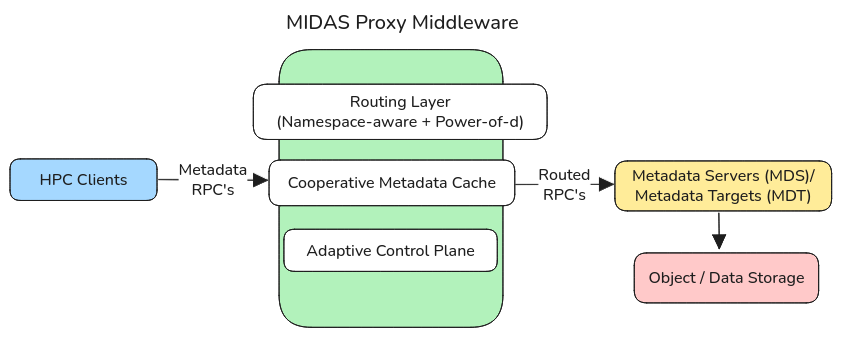}
    \caption{Architecture of \gls{midas}}
    \label{Figure: Architecture}
\end{figure}

As shown in Figure~\ref{Figure: Architecture}, \gls{midas}, deployed as daemon processes, connects to \gls{mds}.
It terminates metadata \gls{rpc}, applies routing and caching policies, if applicable, and then forwards to the chosen \gls{mdt} through \gls{mds}.
During the process of routing and caching evaluation, \gls{midas} uses the \gls{ipc} from \gls{mds} and forwards the call to \gls{mdt} to further write into the \gls{mdt}.

\subsection{Routing Layer}
\label{subsec:routing-layer}
This layer assigns each metadata request to an appropriate \gls{mdt} via the \gls{mds}.
To achieve this, the routing logic combines three complementary mechanisms listed below:
\begin{itemize}
    \item \textbf{Consistent-hashing baseline.} The middleware does not alter existing server behavior; instead, it consults the hash table already maintained by the \gls{mds} (e.g., as in Lustre) via \gls{ipc}. By default, each namespace object like inode, directory, is mapped by consistent hashing to a primary server set, which as a result yields balanced distribution and stable object to server mappings~\cite{LustreWIKI}.
    \item \textbf{Power-of-$d$ refinement.} After consistent hashing baseline, the sample candidate servers $d$, from the consistent-hash mapping, routes the request to the one with the lowest observed load, using server-reported telemetry such as in-flight \gls{rpc} queue length and recent median latency, while CPU utilization is used only as a tie-breaker.
. This preserves placement stability while opportunistically smoothing short term hotspots.
    \item \textbf{Namespace awareness.} When namespace constraints dictate a specific endpoint such as lock ownership, subtree partitioning, the router addresses those constraints and then forwards, else requests remain eligible for the adaptive choice above, consistent with the existing \gls{mds} policy.
\end{itemize}

\subsection{Cooperative Metadata Caching}
 \gls{midas} maintains a cooperative cache of read-mostly metadata entries to reduce the number of \gls{ipc} and reduce the load to \gls{mds}.
To achieve this, \gls{midas} uses the combination of the following approaches:
\begin{itemize}
    \item \textbf{Cache scope:} Safe-to-cache operations include \texttt{lookup}, \texttt{getattr}, and directory reads that do not alter namespace state.
    \item \textbf{Coherence mechanism:} Each cached entry is associated with a server-issued lease or invalidation token when available. For systems with no explicit invalidation, like BeeGFS, \gls{midas} employ adaptive \glspl{ttl}, with conservative defaults to avoid stale results. \gls{midas} serves cached entries only within their validity horizon i.e lease, explicit invalidation, or conservatively chosen \gls{ttl}, when no leases exist, \glspl{ttl} error on freshness (early expiry), not on serving stale metadata. {\gls{midas} guarantees consistency as cached entries are never served past their validity horizon i.e lease, expiry or \gls{ttl}.
    \item \textbf{Cooperation:}} They share cache state via gossip protocol, ensuring that once metadata is fetched, it serves the same entry until cache invalidation or expiry.
\end{itemize}

\subsection{Adaptive Control Plane}
As workload condition varies over time, aggressive re-routing or overly long cache \glspl{ttl} can lead to instability or stale data. \gls{midas} includes an adaptive control plane that monitors system metrics like server load, request latency, cache hit ratio and thus adjusts it's parameters.
Specifically, the control plane regulates the system along three dimensions that are mentioned below:

\begin{itemize}
    \item \textbf{Routing aggressiveness:} Dynamically adjusts $d$ (the number of sampled servers) to balance routing overhead against hotspot mitigation.
    % However, changing the value of d from 2 to 4 was not significant in the simulated request.
    % But still to be in a safe side the d is updated over time.
    \item \textbf{Cache lifetimes:} Tunes \gls{ttl} based on observed invalidation rates and workload read/write mix.
    \item \textbf{Stability:} Uses damped feedback such as moving averages and hysteresis thresholds to prevent oscillations under bursty workloads.
\end{itemize}

\subsection{Working of the Self-Stabilizing Control Loop}\label{sec:control-loop}

The control loop operates under a set of assumptions, guided by design principles, and supported by telemetry for stable adaptation.

For our control loop to be self-stabilizing, we assume:
\begin{enumerate}
    \item Middleware observes per-\gls{mds} instantaneous queue length $L_i$ (in service $+$ waiting) and latency sketches ($\widetilde{p50}_i,\widetilde{p99}_i$), with at most one fast-interval of delay;
    \item Network \gls{rtt} $\leq C$ (pinning interval);
    \item Cache invalidations are observable (leases when available, else notifications or periodic misses).
\end{enumerate}
Since telemetry is observed with \gls{rtt}-scale delays, \gls{midas} uses \gls{ewma} smoothing and hysteresis margins to avoid misrouting due to stale load information.

% We use observable signals only, keeping actuation monotone and bounded with separate timescales (fast routing vs.\ slow caching) while preventing herd behavior via margins, pinning, and an explicit re-route cap.

Every $T_{\mathrm{fast}}{=}250$\,ms, middleware ingests $\{L_i,\widetilde{p50}_i,\widetilde{p99}_i\}$ and maintains \glspl{ewma} \cite{roberts2000gma}:
\[
\hat{x}_t = (1{-}\alpha)\hat{x}_{t-1} {+} \alpha x_t,\quad \alpha{=}0.2.
\]
Every $T_{\mathrm{slow}}{=}30$\,s, it updates per-class cache statistics, the invalidation inter-arrivals being $\Delta t$ and write fraction $W_c$ (\gls{ewma} with $\alpha{=}0.1$) per cache class $c$.

% \paragraph{Targets and pressure}

We regulate imbalance 
and tail latency $\widetilde{p99}$ against targets $(B_{\mathrm{tgt}}, P99_{\mathrm{tgt}})$ via a pressure score measured as
\[
\mathcal{P} = w_1\,[B{-}B_{\mathrm{tgt}}]_+ + w_2\,[\widetilde{p99}{-}P99_{\mathrm{tgt}}]_+,
\qquad [z]_+ \triangleq \max(z,0).
\]

% \paragraph{Knobs and timescales}

\emph{Fast (every $T_{\mathrm{fast}}$):} sampling degree $d\!\in\!\{1,2,3,4\}$; queue margin $\Delta_L\!\in\!\mathbb{N}$ (how many queued requests better a candidate must be than the primary); latency margin $\Delta_t{>}0$ (ms); and a middleware reroute cap $f_{\max}\!\in\![0,f_{\mathrm{cap}}]$ (max fraction of eligible requests steered in a sliding window).\\
\emph{Slow (every $T_{\mathrm{slow}}$):} per-class $\mathrm{\gls{ttl}}_c$, always capped by lease expiry when available.

% \paragraph{Routing rule: JBT with dual margins and pinning}
For a request with feasible set $\mathcal{F}(r)$ and primary $p$, for consistent hashing, sample $S\subseteq\mathcal{F}(r)$ of size $d$.
We \emph{consider} steering only if there exists $j\in S$ such that
\[
\hat{L}_j \le \hat{L}_p - \Delta_L
\quad \text{and} \quad
\widetilde{p50}_j \le \widetilde{p50}_p - \Delta_t .
\]
Among eligible $j$, we choose the one with minimum $\hat{L}_j$, else we keep $p$.
To avoid flapping, we pin the namespace shard, key to its chosen server for at least $C$\,ms before re-evaluation, with $C \ge \text{\gls{rtt}}$ and $C \ge T_{\mathrm{fast}}$.

% \paragraph{Reroute cap (no coordination required)}
Middleware enforces a leaky-bucket cap on steering: in any sliding window of length $W$, no more than an $f_{\max}$ fraction of \emph{eligible} requests may be routed away from their primary. If the bucket is empty, we keep the primary even if margins would allow steering. We keep $f_{\max}\le f_{\mathrm{cap}}$ (e.g., $0.10$--$0.25$) to bound worst-case aggregate steering even if many proxies react similarly.

% \paragraph{Fast-loop actuation (bounded, hysteretic)}
With deadbands $H_\downarrow{<}H_\uparrow$ and hysteresis counters ($K_\uparrow{=}3$, $K_\downarrow{=}8$ fast-intervals), we adjust knobs in single bounded steps:
\[
\begin{aligned}
\text{if } & \mathcal{P} > H_\uparrow \text{ for } K_\uparrow \text{ iters:} \\  
& d \leftarrow \min(d+1,4), \;\;
\Delta_L \leftarrow \max(\Delta_L-1,\, \Delta_L^{\min}); \\[2mm]
\text{if } & \mathcal{P} < H_\downarrow \text{ for } K_\downarrow \text{ iters:} \\  
& d \leftarrow \max(d-1,1), \;\;
\Delta_L \leftarrow \min(\Delta_L+1,\, \Delta_L^{\max});
\end{aligned}
\]
We add small jitter to $\Delta_t$ ($\pm 0.1$\,\gls{rtt}) to avoid lockstep moves across proxies.

% \paragraph{\gls{ttl} via hazard targeting (slow, lease-capped)}

For cache class $c$, estimate invalidation hazard
\[
\hat{h}_c \leftarrow (1{-}\beta)\hat{h}_c + \beta\cdot \frac{1}{\Delta t}, \quad \beta{=}0.1.
\]
Given target stale probability $p^\star$ (e.g., $10^{-4}$), set
\[
\mathrm{\gls{ttl}}_c \;\leftarrow\; \min\!\big(\text{lease\_remaining},\ -\ln(1{-}p^\star)/\hat{h}_c\big),
\]
and multiply by $\gamma{<}1$ (e.g., $0.5$) if $W_c>W_{\mathrm{high}}$. \gls{ttl}s update only on the slow loop and never below a transport floor ($\mathrm{\gls{ttl}}_{\min}\!\ge$ one \gls{rtt}).
\vspace{0.2cm}
\subsubsection{\textbf{Why \gls{midas} self-stabilizes (proof )}}

 Let $V(\hat{L}) \triangleq \sum_i (\hat{L}_i - \bar{L})^2$ with $\bar{L}=\tfrac{1}{m}\sum_i \hat{L}_i$. 

\textbf{Lyapunov Stability \cite{article_lyapunov} perspective:}  
For system stability, we require that every routing action decreases the potential function, i.e., $\Delta V < 0$.

Consider moving one request from $p$ to $j$ (the JBT action) when $\hat{L}_p - \hat{L}_j \ge \Delta_L$. The change in potential is 
\begin{align}
\Delta V &= \big((\hat{L}_p{-}1{-}\bar{L})^2 - (\hat{L}_p{-}\bar{L})^2\big) 
+ \big((\hat{L}_j{+}1{-}\bar{L})^2 - (\hat{L}_j{-}\bar{L})^2\big) \\
&= 2(\hat{L}_j - \hat{L}_p) + 2.
\end{align}

\textbf{Stability condition:} For Lyapunov stability, we need $\Delta V < 0$:
\begin{align}
2(\hat{L}_j - \hat{L}_p) + 2 &< 0 \\
\hat{L}_j - \hat{L}_p &< -1 \\
\hat{L}_p - \hat{L}_j &> 1
\end{align}

If $\Delta_L \ge 2$, then $\hat{L}_p{-}\hat{L}_j \ge 2 > 1 \Rightarrow \Delta V \le -2 < 0$, i.e., each admitted move strictly reduces $V$ and guarantees Lyapunov stability. For a batch of size $m$, $\Delta V = 2m(\hat{L}_j - \hat{L}_p) + 2m^2$, requiring $\hat{L}_p - \hat{L}_j > m$ for a strict decrease (equality yields $\Delta V = 0$).

Therefore, $\Delta_L \ge 2$ is the minimum threshold that ensures convergence to load balance. 

% Because (i) we only steer when \emph{both} margins hold (queue and latency), (ii) we cap steering by $f_{\max}$ per window, and (iii) we pin for $C$\,ms, the aggregate drift of $V$ is negative whenever $\mathcal{P}>H_\uparrow$ and cannot overshoot due to the cap and pinning. As $\mathcal{P}$ re-enters $[H_\downarrow,H_\uparrow]$, hysteresis moves $(d,\Delta_L)$ back toward baseline monotonically. \gls{ttl}s operate on a slower timescale and are capped by leases, so cache policy does not chase fast bursts. Hence the loop returns to and remains within an invariant region 
% \[
% \big\{\,B \le B_{\mathrm{tgt}}{+}\epsilon,\;\; \widetilde{p99}\le P99_{\mathrm{tgt}}{+}\epsilon\,\big\}
% \]
% without oscillation.

\subsection{Algorithm}
The proposed algorithm is given in Algorithm~\ref{alg:midas-control}.

\begin{algorithm}
\caption{\gls{midas} Self-Stabilizing Control for Middleware}
\label{alg:midas-control}
\begin{algorithmic}[1]

    \Require $T_{\mathrm{fast}} = 250\,\text{ms}$, $T_{\mathrm{slow}} = 30\,\text{s}$, $H_\downarrow < H_\uparrow$, $K_\uparrow = 3$, $K_\downarrow = 8$
    \Ensure Stable routing with controlled reroutes

    \State \textbf{Initialize defaults:}
    \State $T_{\mathrm{fast}} \gets 250\,\text{ms}$
    \State $T_{\mathrm{slow}} \gets 30\,\text{s}$
    \State $d \gets 2$ \Comment{$d \in \{1,2,3,4\}$}
    \State $\Delta_L \gets 4$
    \State $\Delta_L^{\min} \gets 2$
    \State $\Delta_L^{\max} \gets 8$ \Comment{requests}
    \State $\Delta_t \gets \text{\gls{rtt}}$
    \State $\mathrm{\gls{ttl}}_{\min} \gets \text{one \gls{rtt}}$
    \State $C \gets 300\,\text{ms}$
    \State $f_{\mathrm{cap}} \gets 0.1$ \Comment{with $f_{\max} \le f_{\mathrm{cap}}$}
    \State $f_{\max} \gets f_{\mathrm{cap}}$
    \State $H_\downarrow \gets 0.02$
    \State $H_\uparrow \gets 0.10$
    \State $K_\uparrow \gets 3$
    \State $K_\downarrow \gets 8$
    \State $w_1 \gets 1$, $w_2 \gets 1$
    \State $\varepsilon \gets 10^{-6}$
    \State $W \gets 1\,\text{s}$ \Comment{reroute window}
    \Statex
    \State \textbf{Convention:} unless stated otherwise, set $w_1 = w_2 = 1$, use $\varepsilon = 10^{-6}$, and reroute window $W = 1\,\text{s}$.
    \Statex

    \State Perform warmup and derive $(B_{\mathrm{tgt}}, P99_{\mathrm{tgt}})$ per \text{\S\ref{sec:targets}}
    \Statex

    \While{every $T_{\mathrm{fast}}$}
        \State Ingest per-\gls{mds} telemetry $\{L_i, \widetilde{p50}_i, \widetilde{p99}_i\}$
        \State Update \glspl{ewma} $\hat{L}_i$, $\widetilde{p50}_i$, $\widetilde{p99}_i$
        \State $B \gets \dfrac{\mathrm{std}(\hat{L})}{\mathrm{mean}(\hat{L}) + \varepsilon}$
        \State $\mathcal{P} \gets w_1\,[B - B_{\mathrm{tgt}}]_+ + w_2\,[\widetilde{p99} - P99_{\mathrm{tgt}}]_+$
        \State Update leaky-bucket tokens enforcing reroute cap $f_{\max}$ over window $W$ (e.g., $1\,\text{s}$)
        \Statex

        \If{$\mathcal{P} > H_\uparrow$ for $K_\uparrow$ consecutive iterations}
            \State $d \gets \min(d + 1, 4)$
            \State $\Delta_L \gets \max(\Delta_L - 1, \Delta_L^{\min})$
        \ElsIf{$\mathcal{P} < H_\downarrow$ for $K_\downarrow$ consecutive iterations}
            \State $d \gets \max(d - 1, 1)$
            \State $\Delta_L \gets \min(\Delta_L + 1, \Delta_L^{\max})$
        \EndIf

        \State Add small middleware jitter to $\Delta_t$ (e.g., $\pm 0.1\,\text{\gls{rtt}}$)
        \Statex

        \ForAll{incoming request $r$}
            \State Let primary $p \gets$ consistent-hash target of $r$
            \State Sample $S \subseteq \mathcal{F}(r)$ of size $d$
            \State $\mathcal{E} \gets \left\{\, j \in S \;\middle|\; \hat{L}_j \le \hat{L}_p - \Delta_L \;\wedge\; \widetilde{p50}_j \le \widetilde{p50}_p - \Delta_t \,\right\}$
            \If{$\mathcal{E} \neq \emptyset$ \textbf{and} reroute tokens are available}
                \State $j^\star \gets \arg\min_{j \in \mathcal{E}} \hat{L}_j$ \Comment{random tie-break on ties}
                \State Route $r$ to $j^\star$
                \State Consume one reroute token
                \State Pin $(\text{key} \rightarrow j^\star)$ for duration $C$
            \Else
                \State Route $r$ to primary $p$
            \EndIf
        \EndFor
    \EndWhile
    \Statex

    \While{every $T_{\mathrm{slow}}$}
        \ForAll{cache class $c$ (backend/namespace/op)}
            \State Update hazard $\hat{h}_c \gets (1 - \beta)\,\hat{h}_c + \beta \cdot (1 / \Delta t)$ with $\beta = 0.1$
            \State Read workload metric $W_c$
            \State $\mathrm{\gls{ttl}}_c \gets \min\!\bigl(\text{\texttt{lease\_remaining}},\, -\ln(1 - p^\star) / \hat{h}_c\bigr)$
            \Comment{$p^\star$ e.g.\ $10^{-4}$}
            \If{$W_c > W_{\mathrm{high}}$}
                \State $\mathrm{\gls{ttl}}_c \gets \max\!\bigl(\mathrm{\gls{ttl}}_{\min},\, \gamma \cdot \mathrm{\gls{ttl}}_c\bigr)$
                \Comment{$\gamma < 1$, e.g.\ $0.5$}
            \EndIf
        \EndFor
    \EndWhile

\end{algorithmic}
\end{algorithm}

\section{Theoretical Model and Results}\label{sec:theory}

\subsection{Power-of-$d$ Choices for Metadata Routing}
We model request routing as a balls-into-bins process, where each incoming metadata request is a ``ball'' and metadata server is a ``bin.'' In the baseline (uniform hashing), each ball is placed into one randomly chosen bin. This yields an expected maximum load of $\frac{\ln M}{\ln \ln M}$ higher than the mean~\cite{azar1999balanced}.

With the \emph{power-of-$d$ choices} scheme, each ball samples $d$ candidate bins and is placed into the one with the smallest current load. The maximum load becomes $\frac{\ln \ln M}{\ln d} + O(1)$ above the mean, exponentially improving queue balance compared to uniform random placement~\cite{mitzenmacher2001power}.

\subsection{Queueing-Theoretic Latency Bounds}
%Follows M/M/1 as the request once routed would be handled by that server and would not be hopped onto next
Under an M/M/1  queueing assumption, the expected latency at server $s_i$ is:
%need to show the wait time and service time too??
\[
E[T_i] = \frac{1}{\mu_i - \lambda_i}, \quad \text{for } \lambda_i < \mu_i.
\]
%expected latency at ith server, this will be calculated accross all server
% ananlyse this across the server as the deviation between this expected latency should be lowered 

% maybe define goal here

  By distributing equal load across the server the maximum $\lambda_i$ can be reduced, lowering both average and tail latency.Building upon the principle that load distribution is key to performance, \gls{midas} improves latency bounds compared to static hashing  by ensuring that $\max_i \lambda_i$ is minimized through adaptive routing.

\subsection{Analytical Insights}
Our analysis depicts three key results:
\begin{enumerate}
    \item \textbf{Load imbalance reduction:} Power-of-$d$ routing reduces the gap between maximum and average load from $\Theta\!\left(\frac{\ln M}{\ln \ln M}\right)$ to $\Theta(\ln \ln M)$, even under bursty arrivals.
    \item \textbf{Tail latency mitigation:} Lower maximum load directly reduces 99th-percentile latency in M/M/1 queues, aligning with our experimental results.
    \item \textbf{Adaptivity requirement:} A feedback-driven adaptation of $d$ and cache \gls{ttl}s are necessary to sustain improvements in dynamic environments.
\end{enumerate}
These theoretical insights motivate the design of \gls{midas}, which integrates namespace-aware power-of-$d$ routing, cooperative caching, and adaptive feedback control into a deployable middleware.

\subsection{Complexity and Overhead Analysis}

MIDAS computational overhead is caused from routing, caching, and adaptive control operations. For each metadata request $r$, at most $d$ candidate servers are sampled by routing layer within the feasible namespace set $F(r)$, resulting in a per-request cost of $O(d)$, where $d \le 4$. Telemetry collection and EWMA smoothing have constant time complexity $O(1)$ per control interval, while the statistics across $m$ metadata servers are updated periodically using control loop with $O(m)$ cost. Similarly middleware keeps track of per-server telemetry tables and cooperative caches storing up to $C$ metadata entries, yielding a total space complexity of $O(m + C)$. Asynchronous invalidation and gossip updates take place with negligible amortized overhead. Altogether, MIDAS has minimal computational and memory overhead, achieving lightweight and scalable behaviour which is favourable for large-scale HPC environments.

\section{Results, Evaluation and Comparison} \label{sec:eval&comparison}
\begin{figure}[h]
    \centering
    \includegraphics[width=\linewidth]{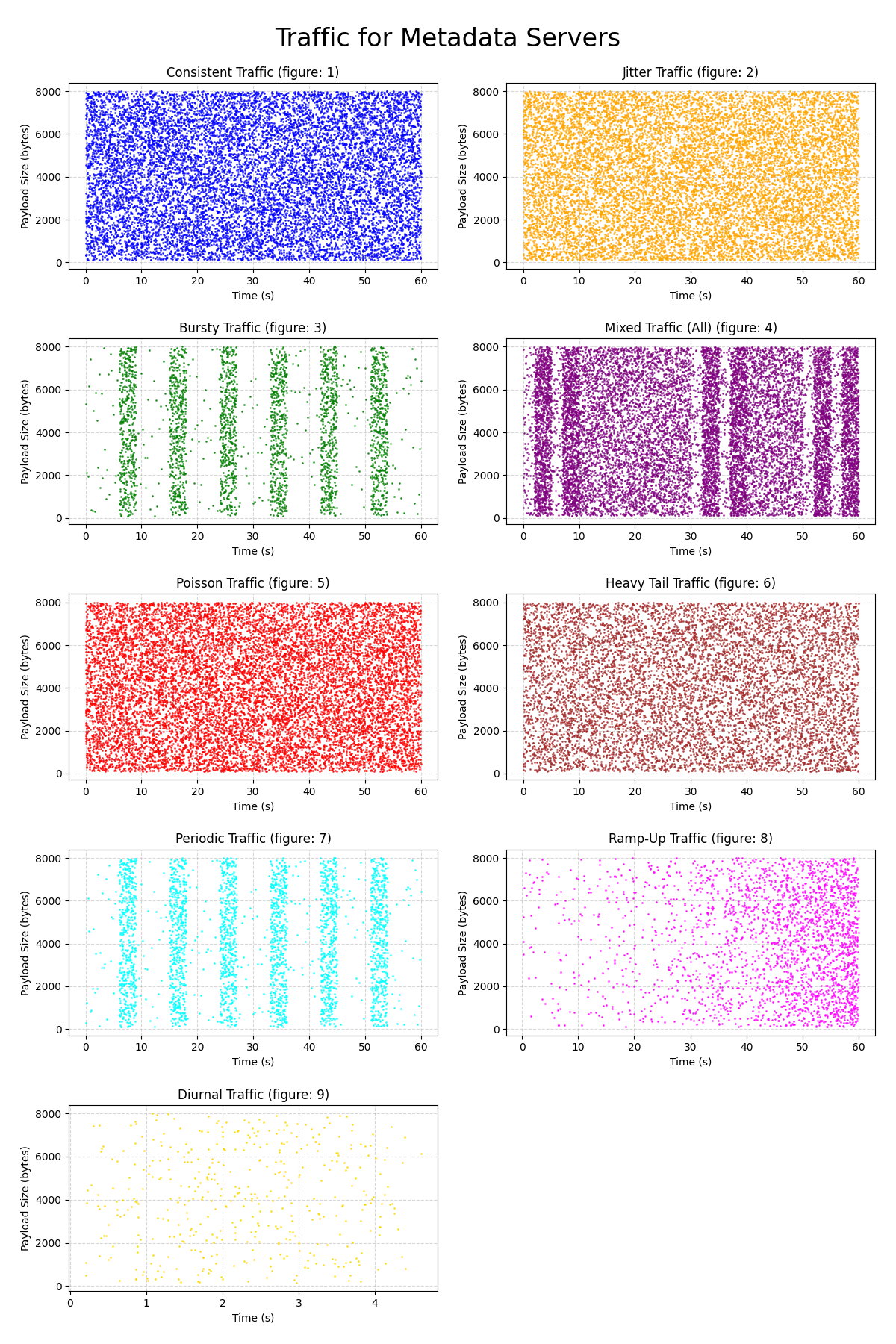}
    \caption{Traffic patterns across different workloads. Each dot corresponds to an \gls{rpc} request.}
    \label{Figure: traffic}
\end{figure}

\begin{figure}[h]
    \centering
    \includegraphics[width=\linewidth]{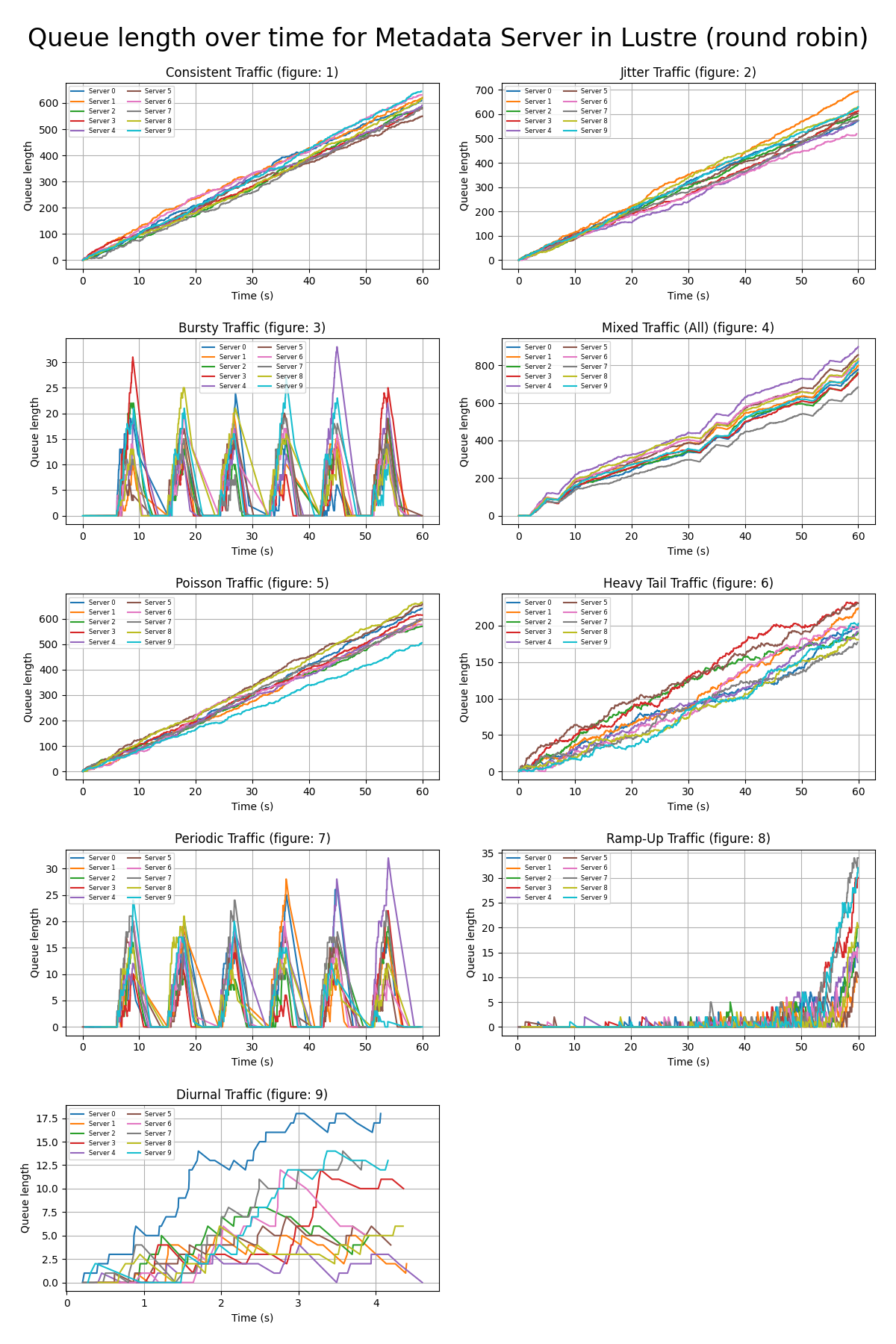}
    \caption{Queue length over time using round-robin distribution (Lustre baseline).}
    \label{Figure: roundrobin}
\end{figure}

\begin{figure}[h]
    \centering
    \includegraphics[width=\linewidth]{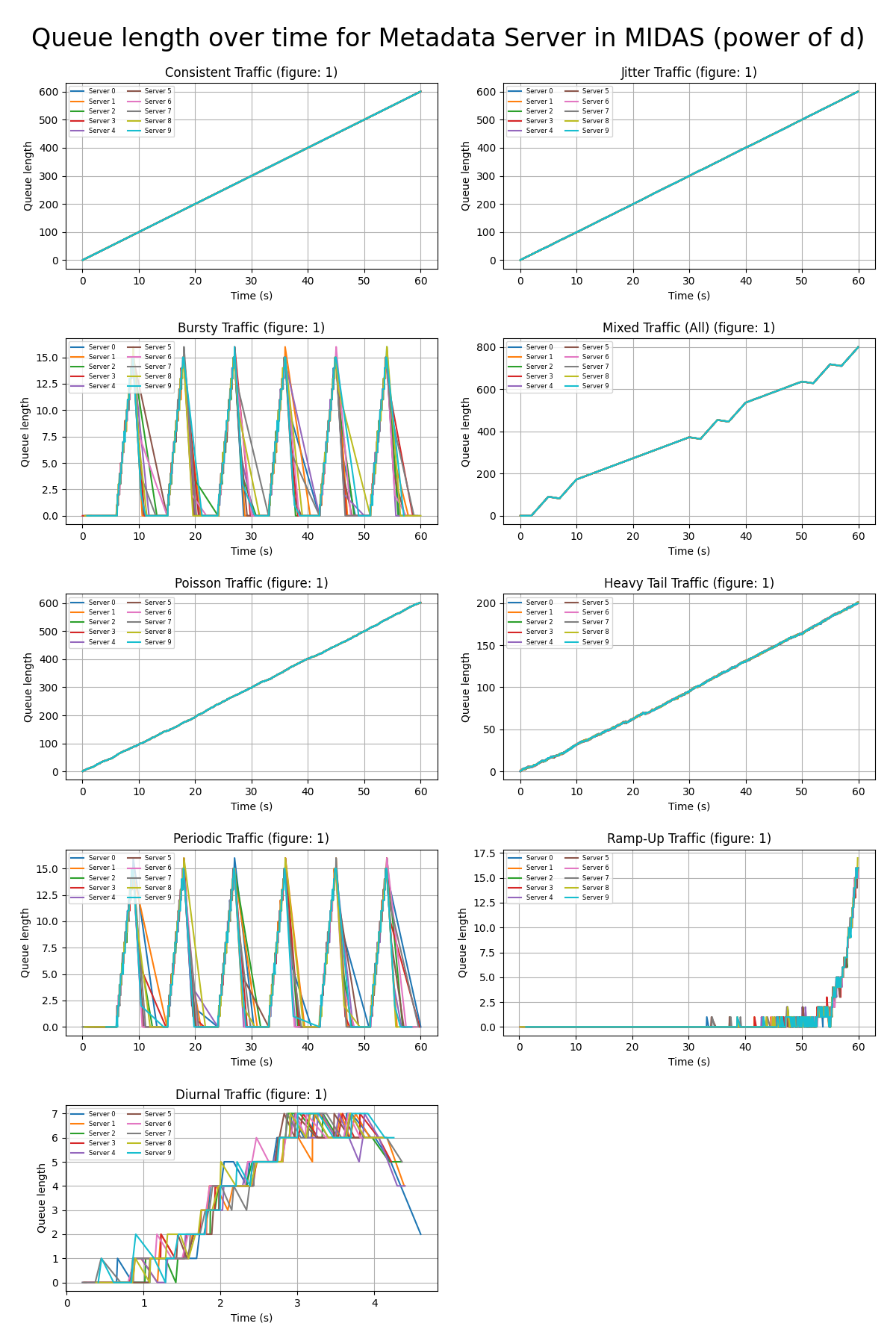}
    \caption{Queue length over time using \gls{midas} distribution (power-of-$d$).}
    \label{Figure:midas}
\end{figure}
\subsection{Assumption}
For comparison with existing methods, we base our analysis on the following assumptions:
\begin{enumerate}
    \item Metadata server is assumed to have access to the complete 'namespace' required for request processing. 
    \item \gls{mdt} requires a constant $100\,\text{ms}$ service time per \gls{rpc} request as a stress bound under overload to magnify queueing; production metadata service times are typically lower, but our conclusions target relative improvements under skew.
\end{enumerate}
These assumptions create a controlled, comparable evaluation environment even, allowing \gls{midas} to highlight adaptive routing and hotspot mitigation.

\subsection{Experimental Setup}
Figure~\ref{Figure: traffic} illustrates the  traffic patterns with which \gls{midas} was tested upon, here each dot (point) represents a \gls{rpc} metadata request.

Two scheduling strategies were evaluated:
\begin{itemize}
    \item \textbf{Lustre (Round-Robin)}: Requests are assigned sequentially across metadata targets. This method is widely used in existing parallel file systems due to its simplicity~\cite{LustreWIKI}.
    \item \textbf{\gls{midas}}: Requests are distributed using the \textit{power-of-$d$} choice algorithm. 
\end{itemize}

\subsection{Results and Performance Analysis}

The queue lengths for both methods were captured under identical traffic workloads as shown in Figure \ref{Figure: roundrobin} and Figure \ref{Figure:midas}. The comparative results highlight the efficiency of \gls{midas} over round-robin:
\begin{itemize}
    \item \textbf{Average queue length}: \gls{midas} reduces the average queue length by approximately $23\%$.
    \item \textbf{Worst-/best-case improvement}: Across workloads, \gls{midas} achieves $50\%$--$80\%$ shorter queues in worst cases.
    \item \textbf{Hotspot mitigation}: Under bursty and periodic workloads, round-robin scheduling produces significant hotspots, with long queues forming on specific servers. \gls{midas} balances requests more effectively, mitigating hotspot issues.
\end{itemize}

We quantify imbalance via \textit{dispersion}, defined as the coefficient of variation of per-server queue length (standard deviation divided by mean) over the run. Beyond queue length reduction, \gls{midas} also improves stability in distributing requests:
\begin{itemize}
    \item Under round-robin scheduling, load imbalance is seen. Dispersion in queue length  distribution range from $20\%$ in light scenarios to as high as $88\%$ under bursty or diurnal traffic.
    \item With \gls{midas}, best-case scenarios achieve perfect uniformity (\textbf{zero dispersion}), while even in challenging traffic patterns, dispersions are limited to around $43\%$.
\end{itemize}

\subsection{Summary}
Overall, the \gls{midas} algorithm consistently delivers:
\begin{enumerate}
    \item Lower queue lengths across all workloads.
    \item Improved fairness and stability in request distribution.
    \item Significant reduction of hotspot issues compared to round-robin scheduling.
\end{enumerate}
These results demonstrate the advantages of applying the power-of-$d$ strategy in metadata server request scheduling, establishing \gls{midas} as a more scalable and balanced alternative to conventional round-robin approaches.

\section{Conclusion} \label{sec:conclusion}

We introduced \gls{midas}, an adaptive middleware for mitigating metadata hotspots. In controlled experiments reflecting \gls{hpc} burst patterns, \gls{midas} combines namespace-aware routing, cooperative caching, and a self-stabilizing control loop, without any client or server changes. \gls{midas} achieves scalability and predictability using approaches like namespace-aware routing, cooperative caching, and a self-stabilizing control loop that doesn't need any modifications to clients or servers.\newline
Proposed middleware ensures stability and demonstrates clear benefits with reduction in mean queue length by 23\% and mitigation of worst-case hotspot issues up to 80\% as compared to Lustre (Round-Robin) scheduling. Our results show that feedback-driven, middleware-based approaches for metadata management is effective. \gls{midas} can lead a way to become self-adaptive metadata service that bridge \gls{hpc} and cloud environments by improving scalability, stability, and performance. These resultsc depends on service-time distribution, backend lease/invalidation availability, and telemetry staleness, which could affect routing accuracy in different production deployments.
Further work focuses on integrating \gls{midas} with petascale production servers, and exploring its applicability to cloud-native storage and orchestration platforms.

\bibliographystyle{IEEEtran}
\bibliography{Bibliography}
\end{document}